\documentclass[apjl]{emulateapj}

\usepackage{graphics}
\shorttitle{BAO Fitting vs Full Modeling}
\shortauthors{SHOJI, JEONG $\&$ KOMATSU}
\begin{document}
\title{
  Extracting Angular Diameter Distance and Expansion Rate of
  the Universe from Two-dimensional Galaxy Power Spectrum 
  at High Redshifts: Baryon Acoustic Oscillation Fitting versus Full Modeling
}
\author{Masatoshi Shoji, Donghui Jeong $\&$ Eiichiro Komatsu}
\affil{Department of Astronomy, University of Texas at Austin, \\
       1 University Station, C1400, Austin, TX, 78712}
\email{mshoji@astro.as.utexas.edu}
\begin{abstract}
 We present a method for extracting the angular diameter distances, $D_A$, and
 the expansion rates, $H$, of the universe from the {\it two-dimensional} 
 Baryon Acoustic Oscillations (BAO) in the galaxy power spectrum. Our
  method builds upon the existing algorithm called the
 ``fit-and-extract'' (FITEX) method, which allows one to extract only
 $D_A^2/H$ from a spherically averaged one-dimensional power
 spectrum. We develop the FITEX-2d method, an extension of the FITEX
 method, to 
 include the two-dimensional  information, which allows us to extract
 $D_A$ and $H$  
 simultaneously. We test the FITEX-2d method using the Millennium Simulation as
 well as simplified Monte Carlo simulations with a bigger volume. The
 BAOs, however, contain only a limited amount of information.
 We show that the full modeling, including the overall shape of the
 power spectrum, yields much better determinations of $D_A$ and $H$,
 hence the dark energy equation of state parameters such as $w_0$ and
 $w_a$,  than 
 the BAO-only analysis by more than a factor of two, provided that non-linear
 effects are under control.
\end{abstract}
\section{Introduction}

Dark energy, discovered via the observed luminosity distances 
out to high-$z$ Type Ia supernovae
\citep{riess/etal:1998,perlmutter/etal:1999},
is the most mysterious element in physics today
\citep[see][for a recent review]{copeland/sami/tsujikawa:2006}.

As dark energy primarily affects the expansion rate of the universe,
one can gain information on the nature of dark energy by measuring 
the cosmological distances as well as the expansion rates of the
universe accurately.\footnote{While dark energy also affects the growth rate of
the amplitude of matter fluctuations, which has been seen in the data
via the so-called Integrated Sachs--Wolfe (ISW) effect 
\citep[e.g.,][]{boughn/crittenden:2004,nolta/etal:2004,afshordi/loh/strauss:2004}, 
we do not discuss the effect on the amplitude of fluctuations in this paper.}

While the cosmic microwave background (CMB) and the Type Ia supernovae
can be used for measuring the angular diameter distance out to $z\simeq 1090$
and the luminosity distances 
out to $z\lesssim 2$, respectively, 
the power spectrum of matter distribution in the universe can be 
used to measure the angular diameter distances {\it as well as} the
expansion rates of the universe out to a wider range of redshifts.

{\it Two} length scales are encoded in the matter power spectrum, $P(k)$
\citep[see, e.g.,][]{weinberg:COS}:
\begin{itemize}
 \item The comoving Hubble horizon size at the matter-radiation equality,
       $r_H(z_{eq})=c/[a(z_{eq})H(z_{eq})]$.
 \item The comoving sound horizon size at the so-called drag epoch
       at which baryons were released from photons, 
       $r_s(z_{drag})=\int_0^{t(z_{drag})}dt~c_s(t)/a(t)$, where
       $c_s(t)=c/\left[\sqrt{3}(1+a(t)3\Omega_b/(4\Omega_\gamma))\right]$
       is the sound speed of photon-baryon fluid.
\end{itemize}
The former determines the overall shape of the power spectrum of dark
matter including the location of the peak of $P(k)$ at $k_{eq}\equiv
1/r_H(z_{eq})$, whereas the latter determines the location of the
baryonic features called the Baryon Acoustic
Oscillations (BAOs). 

These length scales can be predicted from the 5-year data of the
Wilkinson Microwave Anisotropy Probe (WMAP)
\citep{hinshaw/etal:prep,dunkley/etal:prep,komatsu/etal:prep}\footnote{These
predictions assume a flat universe and dark energy being the vacuum
energy. For a non-flat universe with dark energy having a constant
equation of state, $w$, the WMAP 5-year data yield
$k_{eq}=(0.975^{+0.044}_{-0.045})\times 10^{-2}~{\rm 
Mpc}^{-1}$, $r_s(z_{drag})=153.4^{+1.9}_{-2.0}~{\rm Mpc}$,
$z_{eq}=3198^{+145}_{-146}$, and $z_{drag}=1019.8\pm 1.5$.}:
\begin{eqnarray}
 k_{eq} \equiv \frac1{r_H(z_{eq})} &=& (0.968\pm 0.046)\times
  10^{-2}~{\rm Mpc}^{-1},\\
 r_s(z_{drag}) &=& 153.3 \pm 2.0~{\rm Mpc},
\end{eqnarray}
and
\begin{equation}
 z_{eq}=3176^{+151}_{-150},\qquad z_{drag}=1020.5\pm 1.6.
\end{equation}
These lengths can be used as the ``standard rulers,'' which give us
the angular diameter distances as well as the expansion rates of the
universe
\citep{seo/eisenstein:2003,blake/glazebrook:2003,hu/haiman:2003}.\footnote{The matter power spectrum also contains the third
distance scale, the Silk damping scale, which can also be used as the
standard ruler. The Silk damping scale is the smallest of these three
distance scales, and its effect (i.e., the suppression of power below
the Silk damping scale) is not as prominent as the effects of the other
two distance scales. Nevertheless, the Silk damping must be taken into
account when we model the full shape of the power spectrum.}

We, as observers who measure the angular and redshift distribution of
galaxies, can measure {\it four} distance ratios given by 
\begin{eqnarray}
 \theta_{eq}(z) &=& \frac{r_H(z_{eq})}{(1+z)D_A(z)}
=\frac{1}{k_{eq}(1+z)D_A(z)}, \\
 \theta_{s}(z) &=& \frac{r_s(z_{drag})}{(1+z)D_A(z)}, \\
 \delta z_{eq}(z) &=& \frac{r_H(z_{eq})H(z)}{c}
= \frac{H(z)}{k_{eq}c}, \\
 \delta z_{s}(z) &=& \frac{r_s(z_{drag})H(z)}{c},
\end{eqnarray}
where $D_A(z)$ is the proper (i.e., not comoving) angular diameter distance.
We measure $\theta_{eq}(z)$ and $\theta_s(z)$ by comparing the
predicted lengths with the  
corresponding observed lengths perpendicular to the line of sight,
 and $\delta z_{eq}(z)$ and $\delta z_s(z)$ from the lengths parallel to the
 line of sight.\footnote{The measured power spectrum in redshift space is a
function of the wavenumber parallel to the line of sight, $k_\parallel$, and
that perpendicular to the line of sight, $k_\perp$, i.e.,
$P=P(k_\parallel,k_\perp)$.
The angular observables, $\theta_{eq}$ and $\theta_s$, are measured from
$k_\perp$, while the line-of-sight observables, $\delta z_{eq}$ and
$\delta z_s$, are measured from $k_\parallel$.}

The BAOs have been detected in 
the current galaxy redshift survey data from the Sloan Digital Sky
Survey (SDSS) and the 
Two-degree Field Galaxy Redshift Survey (2dFGRS) 
\citep{eisenstein/etal:2005,cole/etal:2005,hutsi:2006,percival/etal:2007c}. However, the current
data are not yet sensitive enough to yield $D_A(z)$ and $H(z)$
separately \citep{okumura/etal:prep}; thus, one can only determine a
combined distance scale ratio from the spherically averaged power
spectrum. Since two spatial dimensions are available on the sky and one
dimension is available along the line of sight, one can measure 
\begin{equation}
 \left[\theta_s^2(z) \delta z_s(z)\right]^{1/3} 
= \frac{r_s(z_{drag})}{[(1+z)^2D_A^2(z)c/H(z)]^{1/3}}.
\end{equation}
\citet{eisenstein/etal:2005} have measured this quantity at $z=0.35$
from the SDSS Luminous Red Galaxies (LRG), and
\citet{percival/etal:2007c} have extended their analysis to include more
data from the SDSS LRG, as well as the SDSS main galaxy samples and the
2dFGRS galaxies at $z=0.2$. 

\citet{komatsu/etal:prep} have combined
these measurements with the CMB distance ratios determined from the WMAP
5-year data, the ``WMAP distance priors,'' to obtain the constraints on
dark energy properties. The analysis performed in
\citet{komatsu/etal:prep} is a proto-type of what one can do in the
future. It is clear that we can gain more information if we can
measure $D_A(z)$ and $H(z)$ simultaneously at various redshifts. 
Therefore, in the future we should be able to perform a much more
sensitive test of dark energy properties by combining 
$D_A(z)$ and $H(z)$ from the future galaxy survey data, and the CMB
distance priors from the future CMB experiments such as Planck.

Moreover, the BAOs capture only a part of information encoded in the
shape of $P(k)$. One would miss another baryonic feature, the Silk
damping scale, by only measuring BAOs. A more serious drawback is that
one would miss the other prominent standard ruler, $k_{eq}$, completely, by only
measuring BAOs. 

Nevertheless, there is one major advantage of using BAOs: the phases (not
the amplitude) of
BAOs are less sensitive to the distortion of the shape of $P(k)$ due to
non-linear matter clustering, non-linear galaxy bias, or non-linear
redshift space distortion
\citep{seo/eisenstein:2005,eisenstein/seo:2007,nishimichi/etal:2007,smith/scoccimarro/sheth:2008,angulo/etal:2008,sanchez/baugh/angulo:prep,seo/etal:2008}. As
a result, many 
studies have 
focused on developing various ways to extract the distance information
from BAOs. 

Most of the previous work focused only on extracting the BAOs
from the spherically averaged $P(k)$ (which gives $D_A^2/H$)
\citep[e.g.,][]{percival/etal:2007c}. 
\citet{yamamoto/etal:2005} have studied the monopole and quadrupole
moments in the galaxy 
power spectrum and their implications for determinations of the dark
energy equation of state parameter, $w$, and concluded that even
in the worst case scenario (i.e., absence of the BAOs feature on the observed
power spectrum), galaxy survey can still provide useful limits on
$w$ from  a combination of the monopole and quadrupole power spectra.
Recently, 
\citet{padmanabhan/white:prep} have explored an extraction of the
quadrupole moment of the two-dimensional power spectrum,
$P(k,\mu)$, which gives a different distance combination,
$D_AH$. 

In this paper, we shall develop a method for extracting $D_A$ and $H$
simultaneously from the two-dimensional BAOs. 
Since we do not use spherical averaging or
truncate the Legendre expansion of BAOs at arbitrary orders, our method uses
more information than most of the previous methods. To our knowledge,
the full two-dimensional extraction of $D_A$ and $H$ from BAOs has been explored
only by \citet{wagner/muller/steinmetz:prep}. 

This paper is organized as follows. In \S~\ref{sec:fitex} we give a 
brief account of the original one-dimensional ``fit-and-extract''
(FITEX) method, which was developed by
\citet{koehler/schuecker/gebhardt:2007} for extracting BAOs from a
spherically averaged one-dimensional $P(k)$. We then extend this method to the
two-dimensional FITEX-2d method by including the full two-dimensional
information without spherical averaging. 
In \S~\ref{sec:montecarlo} we extract $D_A$ and $H$ from simulated noisy
data using the FITEX-2d method, and show that the FITEX-2d yields
unbiased estimates of $D_A$ and $H$. 
In \S~\ref{sec:millennium} we repeat the same analysis for 
a more realistic simulation, using the Millennium Simulation
\citep{springel/etal:2005}. 
In \S~\ref{sec:darkenergy} we propagate errors in $H(z)$ and $D_A(z)$ to 
those in the dark energy equation of state with the parametrization of 
$w(z)=w_0+w_az/(1+z)$. We conclude in \S~\ref{sec:conclusion}.

Throughout this paper we shall use the cosmological parameters given by
$\Omega_m=0.277$, $\Omega_\Lambda=0.723$, $\Omega_b=0.0459$, $n_s=0.962$,
and $h=0.702$ \citep{dunkley/etal:prep,komatsu/etal:prep}, 
which are the maximum likelihood values inferred from 
the WMAP 5-year data \citep{hinshaw/etal:prep} 
combined with the current BAO data \citep{percival/etal:2007c}
and Type Ia supernova data \citep{kowalski/etal:prep}.

\section{FITEX-2d: Methodology}
\label{sec:fitex}
We develop a method for extracting $D_A$ and $H$ simultaneously from
the two-dimensional BAOs  without
spherical averaging. 

Our method builds upon the existing ``fit-and-extract'' (FITEX) method
developed by \citet{koehler/schuecker/gebhardt:2007} for extracting
$D_A^2/H$ from a spherically averaged, one-dimensional $P(k)$. 
The FITEX method extracts BAOs by fitting and removing the
non-oscillatory part of $P(k)$, which leaves only the oscillatory
component, i.e., BAOs. 
\citet{koehler/schuecker/gebhardt:2007} model
the non-oscillatory, smooth part by the following functional form:
\begin{equation} 
P_{smooth}^{1d}(k)=
 \left[\frac{A}{1+Bk^{\delta}}e^{({k/k_1})^{\alpha}}\right]^2 k^{n_s},
\label{eq:Psmooth1d}
\end{equation}
where $n_s$ is the primordial tilt, while $A$, $B$, $\delta$, $k_1$, and
$\alpha$ are free parameters. \citet{koehler/schuecker/gebhardt:2007}
have shown that this function is flexible enough to fit out the smooth
part of the spherically averaged $P(k)$ measured from the Hubble Volume
Simulation \citep{evrard/etal:2002}. They have tested the FITEX method
particularly for a large scale, $k<0.3~h~{\rm Mpc}^{-1}$, at high
redshifts, $1.9<z<3.8$, that are relevant to the Hobby Eberly Dark
Energy Experiment \citep[HETDEX;][]{hill/etal:2004}.

We make a simple extension of the one-dimensional FITEX method by
including angular dependence. We model the two-dimensional smooth power
spectrum by
\begin{eqnarray}
\nonumber
& & P_{smooth}^{2d}(k,\mu) = P_{smooth}^{1d}(k) \\
&\times& \left[1 +
	  g^{(2)}(k)P_2(\mu)+g^{(4)}(k)P_4(\mu)+g^{(6)}(k)P_6(\mu)\right],
\label{eq:Psmooth2d}
\end{eqnarray} 
where $\mu$ is the cosine of the angle $\theta$ between $\mathbf{k}$ 
and the line of sight, i.e., $\mu=\cos\theta$ and
$\tan\theta=k_\perp/k_\parallel$. Therefore, $\mu=0$ and $\mu=1$ for
$k_\parallel=0$ and $k_\perp=0$, respectively. 

Here, 
$P_l(\mu)$ is the Legendre polynomials:
\begin{eqnarray}
 P_2(\mu) &=& \frac12\left(3\mu^2-1\right),\\
 P_4(\mu) &=& \frac18\left(35\mu^4-30\mu^2+3\right),\\
 P_6(\mu) &=& \frac1{16}\left(231\mu^6-315\mu^4+105\mu^2-5\right).
\end{eqnarray}
The odd multipoles must vanish by symmetry.
One may include $l\ge 8$ if necessary, but we find it sufficient to
include the terms only up to $l=6$.

Finally, $g^{(l)}(k)$ is given by the 6th-order polynomials
with only even powers of $k$:
\begin{equation}
 g^{(l)}(k) = a_0^{(l)} + a_2^{(l)} k^2 + a_4^{(l)} k^4 + a_6^{(l)} k^6,
\label{eq:gl}
\end{equation}
where all of $a_i^{(l)}$'s are varied simultaneously for each $l$.
The odd powers must vanish because they are not analytic in
$\mathbf{k}$ \citep{weinberg:COS}. We include the terms only up to $k^6$, as we include the
multipoles up to $l=6$. If, for instance, $l=8$ is included, then $k^8$
may also be included for consistency.

Aside from the primordial tilt, $n_s$, the FITEX-2d contains 17 free
parameters (5 for $P_{smooth}^{1d}(k)$ plus $4\times 3=12$ for the angular
dependence). While it may sound like many, the number of data points
available on the {\it two-dimensional} power spectrum is usually much
larger, and thus our fit is well behaved.

It may be instructive to use the conventional model for the redshift
space power spectrum to show what these
parameters are supposed to 
capture. The leading order angular distortion is
given by the so-called Kaiser effect, which arises from coherent
converging velocity flow toward the linear overdensity region
\citep{kaiser:1987}. 
The linear Kaiser power spectrum is given by
\begin{eqnarray}
\nonumber
 P_{kaiser}^{linear}(k,\mu) 
&=& b_1^2(1+2\beta\mu^2+\beta^2\mu^4)P^{linear}(k)\\
\nonumber
&=& b_1^2\left[\left(1+\frac23\beta+\frac15\beta^2\right)\right.\\
\nonumber
&+& \frac{4}{3}\beta\left(1+\frac37\beta\right)P_2(\mu)\\
&+& \left.\frac{8}{35}\beta^2P_4(\mu)\right]P^{linear}(k),
\end{eqnarray}
where $\beta\equiv f/b_1$ is a $k$-independent function that depends on
the linear galaxy bias, $b_1$, and the cosmological 
parameters (mainly $\Omega_m$) via
\begin{equation}
 f\equiv \frac{d\ln D}{d\ln a},
\end{equation}
where $D$ is the growth factor of linear density fluctuations.
We therefore find
\begin{eqnarray}
 a_0^{(0)} &=& 1\\
 a_0^{(2)} &=&\frac{\frac{4}{3}\beta\left(1+\frac37\beta\right)}
{1+\frac23\beta+\frac15\beta^2}, \\
 a_0^{(4)} &=&\frac{\frac{8}{35}\beta^2}
{1+\frac23\beta+\frac15\beta^2},
\end{eqnarray}
and the other terms are zero.

Another example is the so-called Finger-of-God (FoG) effect,
which arises from random motion within virialized halos. When the
distribution of the pairwise peculiar velocity within a halo is given by
an exponential distribution with the velocity dispersion $\sigma_v^2$
\citep{peebles:1976,davis/peebles:1983}, one finds
\citep{ballinger/peacock/heavens:1996}
\begin{equation}
 P_{FoG}(k,\mu) = \frac{P^{linear}_{kaiser}(k,\mu)}{1+f^2k^2\mu^2\sigma_v^2}.
\end{equation}
While the FoG yields many terms when expanded into the Legendre polynomials,
 it is still a good approximation to truncate the expansion
at $l=6$ if $k$ is sufficiently smaller than $1/\sigma_v$.
Note that the FoG effect yields terms in the form of powers of $(k\mu)^2$;
thus, it makes sense to use the same number for the maximum power of $k$
(see Eq.~(\ref{eq:gl}))
and the maximum multipole 
(see Eq.~(\ref{eq:Psmooth2d}))
of the FITEX-2d fitting function.

In general, neither of these two expressions are adequate.
The linear Kaiser formula is valid only on very large scales, while the
exponential FoG formula is valid only on very small scales.
At the intermediate scales we find more complicated expressions from,
e.g., the 3rd-order perturbation theory
\citep{heavens/matarrese/verde:1998}. To account for these complications 
we have included $k$-dependent coefficients for the Legendre
polynomials.

\begin{figure}[t]
\centering
\rotatebox{0}{%
  \includegraphics[width=6.5cm]{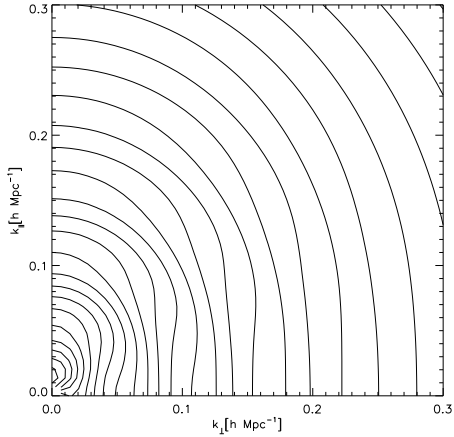}
}%
\caption{%
Illustration of the FITEX-2d method. This figure shows 
an anisotropic non-linear galaxy power spectrum before we apply FITEX-2d. 
The contours show $\ln[P(k_\parallel,k_\perp)]$ at $z=2$, where
 we have computed $P(k_\parallel,k_\perp)$ from
 Eq.~(\ref{eq:pkred}). Anisotropic distribution of power due to 
 redshift space distortion is apparent.
}%
\label{fig1}
\end{figure}

\begin{figure}[t]
\centering
\rotatebox{0}{%
  \includegraphics[width=6.5cm]{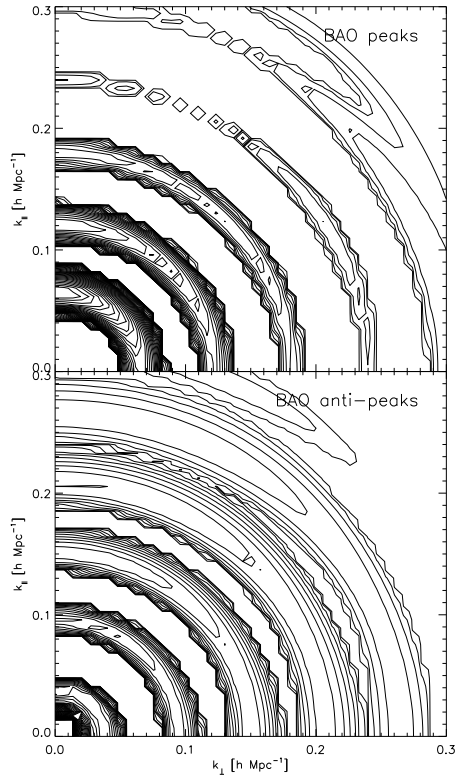}
}%
\caption{%
Illustration of the FITEX-2d method. This figure shows 
the power spectrum shown in Fig.~\ref{fig1} minus the best-fitting
 two-dimensional smooth spectrum,
 $P_{smooth}^{2d}(k_\parallel,k_\perp)$, given by
 Eq.~(\ref{eq:Psmooth2d}).
The structure of BAOs, i.e., the oscillatory feature, is now apparent.
The FITEX-2d method recovers the isotropic distribution of the BAO phases
 successfully, which makes it possible to use the distribution of the phases
 for measuring $D_A$ and $H$ simultaneously.
 ({\it Top}) Positive BAO peaks. ({\it Bottom}) Negative BAO peaks (troughs).
}%
\label{fig2}
\end{figure}

In Figure~\ref{fig1} and \ref{fig2} we show the performance of
$P_{smooth}^{2d}(k,\mu)$. In Figure~\ref{fig1} we show a simple
analytical model\footnote{This model is admittedly too simple to be
realistic. We shall test the FITEX-2d method in a more realistic
setting using the Millennium Simulation in \S~\ref{sec:millennium}.} for the non-linear galaxy power spectrum in redshift space
given by 
\begin{eqnarray}
\nonumber
 P_g(k,\mu) &=& b_1^2\left[P_{\delta\delta}(k) + 2\beta\mu^2P_{\delta\theta}(k)
+\beta^2\mu^4P_{\theta\theta}(k)\right]\\
& &\times\frac1{1+f^2k^2\mu^2\tilde{\sigma}_v^2},
\label{eq:pkred}
\end{eqnarray}
where $P_{\delta\delta}(k)$, $P_{\delta\theta}(k)$, 
and $P_{\theta\theta}(k)$ are the density-density, density-velocity, and 
velocity-velocity power spectra computed from the 3rd-order perturbation
theory, and they are given by 
Eq.~(63), (64), and (65) in \citet{scoccimarro:2004}, respectively. 
This form is similar to Eq.~(71) of \citet{scoccimarro:2004}, but we
have replaced $\exp(-f^2k^2\mu^2\sigma_v^2)$ and $f$ in his formula by 
$1/(1+f^2k^2\mu^2\tilde{\sigma}_v^2)$ and $\beta$, respectively,
where $\tilde{\sigma}_v^2\equiv 0.6\sigma_v^2$ is
the 1-d peculiar velocity dispersion with
 an empirical fudge factor of $0.6$ calibrated off our simulations
 presented in \citet{jeong/komatsu:2006}. Here, $\sigma_v^2$ is given by
\begin{equation}
 \sigma_v^2\equiv
  \frac13\int\frac{d^3k}{(2\pi)^3}\frac{P^{linear}(k)}{k^2}
=
\frac13\int\frac{dk}{2\pi^2}P^{linear}(k).
\label{eq:sigmav2}
\end{equation}

We chose $z=2$ and $b_1=2.5$. 
The contour of power spectrum is anisotropic in Fig.~\ref{fig1} due to the
redshift space distortion; however, we recover isotropy after
subtracting the best-fitting $P_{smooth}^{2d}(k,\mu)$ from the
anisotropic data (see Fig.~\ref{fig2}). We see that the BAOs have been
extracted successfully, with isotropy of the oscillation phases
recovered well.


\section{Extraction of $D_A$ and $H$ from noisy data: FITEX-2d vs Full Modeling}
\label{sec:montecarlo}
In \S~\ref{sec:err_fitex2d} we show how well we can estimate $D_A$ and
$H$ from the 
two-dimensional BAOs extracted from noisy data using the FITEX-2d
method. In \S~\ref{sec:err_full} we compare the BAO results to the accuracy
one would obtain from 
the full modeling of $P(k,\mu)$, including the overall
shape. In other words, for the former (BAOs) we only use $\theta_s$ and
$\delta z_s$ for measuring $D_A$ and $H$, while for the latter (full
modeling) we can use $\theta_s$, $\delta z_s$, $\theta_{eq}$, $\delta
z_{eq}$, as well as the Silk damping scale for measuring $D_A$ and $H$,
provided that non-linear effects (non-linear matter clustering,
non-linear redshift space distortion, and non-linear bias) are under
control. 

Note that the treatment of non-linear effects in this section is too
simple to be realistic. For a more realistic treatment we shall use the
galaxy power spectrum from the Millennium Simulation \citep{springel/etal:2005} in
\S~\ref{sec:millennium}. 
\subsection{FITEX-2d}
\label{sec:err_fitex2d}
To estimate errors in $D_A$ and $H$ from the FITEX-2d method, we use 
simple Monte Carlo simulations. 

For the underlying spectrum we use the same data as shown in
Fig.~\ref{fig1}, which includes a simplified modeling of non-linear
matter clustering and non-linear redshift space distortion as given by
Eq.~(\ref{eq:pkred}). As for the galaxy bias, we use a
linear bias with $b_1=2.5$.

Once the underlying spectrum is specified, it is straightforward to
compute the errors in $P_g(k_\parallel,k_\perp)$, $\sigma_{P_g}$, provided that the
distribution of $P_g(k_\parallel,k_\perp)$ is a Gaussian. We use the
standard formula that includes sampling variance as well as shot noise
\citep[see, e.g.,][]{jeong/komatsu:2008}
\begin{equation}
\frac{\sigma_{P_g}(k_\parallel,k_\perp)}{P_g(k_\parallel,k_\perp)} = 2\pi \sqrt{\frac1{V_{survey}k_{\perp} \Delta
 k_{\perp} \Delta k_{\parallel}}}\frac{1+n_gP_g(k_\parallel,k_\perp)}{n_gP_g(k_\parallel,k_\perp)},
\label{eq:sigP}
\end{equation}
where $n_g$ is the number density of galaxies, $V_{survey}$ is the
survey volume, $\Delta k_\perp$ and $\Delta k_\parallel$ are the
fundamental wavenumbers, i.e., the resolution in $k_\perp$ and
$k_\parallel$. We take these to be $\Delta k_\parallel=\Delta k_\perp=(2\pi)/V_{survey}^{1/3}$.

We use $\sigma_{P_g}$ from Eq.~(\ref{eq:sigP}) to calculate the r.m.s. error
in $P_g(k_\parallel,k_\perp)$, and generate 1000 Monte Carlo
realizations. We then apply the FITEX-2d method to remove the smooth
component from each realization to extract BAOs. For each 
realization, we measure $D_A$ and $H$ 
simultaneously by fitting the phases of extracted two dimensional 
BAOs to those of the reference BAOs extracted from either (i) 
the linear power spectrum, or (ii) the non-linear power spectrum given
by Eq.~(\ref{eq:pkred}), 
with known $D_{A,ref}$ and $H_{ref}$. 
(Later we find that using the linear spectrum as the reference BAO yields the
biased estimates of $D_{A,ref}$ and $H_{ref}$.) 
We use a simplex downhill method 
for $\chi^2$-minimization in the two-dimensional 
parameter space. 
The number of free parameters for this analysis is two, i.e., $D_A$ and
$H$, and we do not include the amplitude in the fit.
We have checked that including the amplitude does not change 
the results very much, as the amplitude and the phases of BAOs are
nearly uncorrelated (see Appendix~\ref{sec:cc_no_beta} for more
details). This is true in both real and redshift space. 
When we apply FITEX-2d to the simulated data, we perform
a fit out to $k_{max}=0.40~h~{\rm Mpc}^{-1}$.

We choose the survey parameters, $V_{survey}$, $z$, and $n_g$, such that
they roughly match those expected for the Hobby-Eberly Dark Energy
Experiment (HETDEX) \citep{hill/etal:2004}: $N_g=0.755\times10^6$,
and $1.9\le z\le 3.5$ with the
sky coverage of 420~deg$^2$, which yields $V_{survey}\simeq 3.0~h^{-3}~{\rm
Gpc}^3$.\footnote{The HETDEX is expected to detect 0.755 million Lyman-$\alpha$
emitting galaxies between $1.9\le z\le 3.5$ over 420~deg$^2$ in 3 years
of observations on the Hobby-Eberly Telescope.}

We find that, when the phases extracted by FITEX-2d are compared with
the reference BAOs extracted from the linear power spectrum, 
 the best-fitting values of $D_A$ and $H$ averaged over 1000
simulations disagree with the underlying, ``true'' values by 0.05\% and 0.63\%
for $D_A$ and $H$, respectively, due to the phase shift
of BAOs caused by non-linearities (including non-linear redshift space
distortion). This result extends the previous study by 
\citet{nishimichi/etal:2007}, who studied a
spherically averaged 1-d power spectrum and found that the bias 
was less than 1\% in $(D_A^2H^{-1})^{1/3}$. 

On the other hand, when the phases are compared with 
the reference BAOs extracted from the {\it non-linear} power spectrum
(Eq.~(\ref{eq:pkred})), the best-fitting values of $D_A$ and $H$ agree
with the true values 
to well within the Monte Carlo sampling error; thus, 
we confirm that the FITEX-2d method yields unbiased estimates of $D_A$
and $H$. 

\begin{figure}[t]
\centering
\rotatebox{0}{%
  \includegraphics[width=7.5cm]{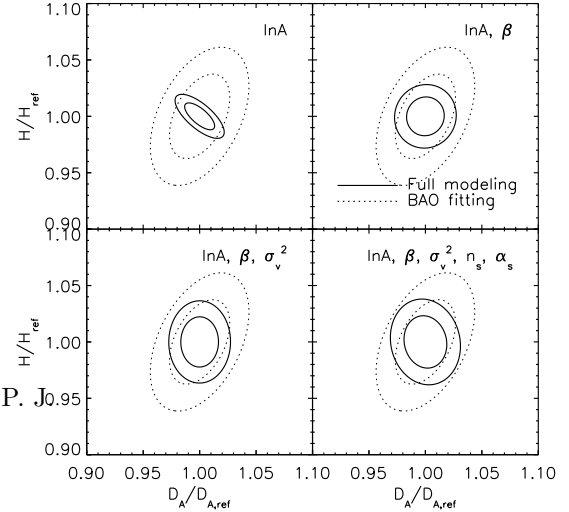}
}%
\caption{%
Accuracy of $D_A$ and $H$ extracted from BAOs with the FITEX-2d method
 applied to simulated Monte Carlo realizations that approximate the
 HETDEX survey (the larger, dotted contours; see \S~\ref{sec:err_fitex2d}).
The best-fitting values of $D_A$ and $H$ agree with the true values;
 thus, the FITEX-2d method yields unbiased estimates of $D_A$ and $H$.
 The solid contours show $D_A$ and $H$ from the full modeling,
 including the overall shape of the power spectrum, with various
 parameters marginalized over. (Note that the BAO-only contours are
 unaffected by the marginalization.)
 For this we have
 used the Fisher matrix forecast (see \S~\ref{sec:err_full}).
The inner and outer ellipses show $68\%$ and
 $95\%$ C.L., respectively.
({\it Top Left}) the full modeling Fisher matrix is marginalized over
 the overall amplitude, $\ln A$,
({\it Top Right}) marginalized over $\ln A$ and the linear redshift
 distortion parameter, $\beta$,
({\it Bottom Left}) marginalized over $\ln A$, $\beta$, and the velocity
 dispersion in the FoG factor, $\tilde{\sigma}^2_v$,
({\it Bottom Right}) marginalized over $\ln A$, $\beta$,
 $\tilde{\sigma}^2_v$, and the shape of the initial power spectrum,
$n_s$ and $\alpha_s$. 
}%
\label{fig3}
\end{figure}

In Figure~\ref{fig3} we show the projected error ellipses on 
 $D_A$ and $H$ from the BAOs extracted with the FITEX-2d 
(larger, dotted contours; same in all four panels).  
We find 1.8\% and 2.5\% errors on
$D_A$ and $H$, respectively, with the cross-correlation coefficient of
$r=0.44$, from the Monte Carlo simulations.
For the same survey parameters, the BAO Fisher matrix proposed by
\citet{seo/eisenstein:2007} yields 1.5\% and 2.5\% errors on
$D_A$ and $H$, respectively, with $r=0.41$. Therefore, we conclude that
the FITEX-2d method yields the results that nearly saturate the Fisher
 matrix bound, i.e., it is nearly an optimal method in a sense that it
 can yield the smallest errorbars one can obtain with the BAO-only analysis.

\subsection{Full Modeling}
\label{sec:err_full}
To calculate the errors in $D_A$ and $H$ expected from the full modeling
of the two-dimensional galaxy power spectrum,
$P_g(k,\mu)$, we use the Fisher matrix given by
\citep[see, e.g.,][]{eisenstein/hu/tegmark:1999,seo/eisenstein:2003}

\begin{equation}
F_{ij}=\int_0^{k_{max}}\frac{4\pi k^2dk}{(2\pi)^3}
\int_0^1d\mu\frac{\partial\ln P_g(k,\mu)}{\partial\theta_i}
\frac{\partial\ln P_g(k,\mu)}{\partial\theta_j}w(k,\mu),
\label{eq:fisher_fm}
\end{equation}

where $\theta_i=(\ln D_A,\ln H,\ln A,\beta,\tilde{\sigma}^2_v,n_s,\alpha_s)$ for $i=1$, 2,...,7, respectively, $k_{max}\!=\!0.40~h~{\rm Mpc}^{-1}$, where $A$ is the
overall amplitude of the power spectrum, $\beta$ is the linear redshift
distortion parameter, $\tilde{\sigma}^2_v$ is the calibrated 1-d
velocity dispersion (see Eq.~(\ref{eq:pkred})), 
and $n_s$ and $\alpha_s$ describe
the shape of the initial (primordial) power spectrum:
\begin{equation}
 P_{ini}(k)\propto k^{n_s+\frac12\alpha_s\ln(k/k_{pivot})}.
\end{equation} 

\begin{figure}[t]
\centering
\rotatebox{0}{%
  \includegraphics[width=7.5cm]{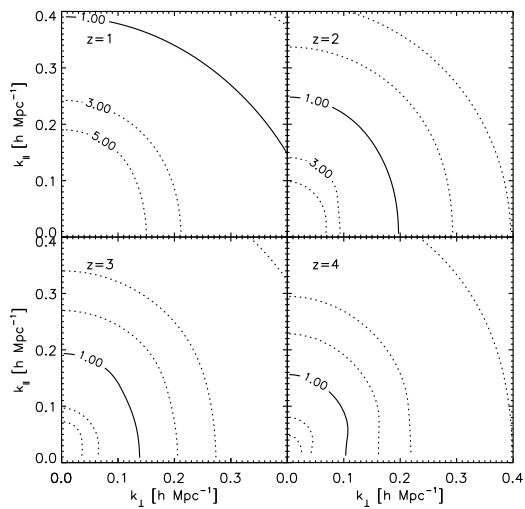}
}%
\caption{%
The galaxy power spectrum times the number density of galaxies,
 $n_gP_g(k,\mu)$, where the number of the galaxies is fixed
for each redshift bin to $N_g=0.755\times10^6$, and
 $P_g(k,\mu)$ is 
 computed from Eq.~(\ref{eq:pkred}) with $b_1=2.5$.
 The shot noise dominates the error budget when $n_gP_g(k,\mu)<1$.
Contour values are [0.1, 0.3, 0.5, 1.0, 3.0, 5.0].
 ({\it Top Left}) $z=1$,
 ({\it Top Right}) $z=2$, 
 ({\it Bottom Left}) $z=3$,
 ({\it Bottom Right}) $z=4$. 
}%
\label{fig4}
\end{figure}

Here, the weight function, $w(k,\mu)$, is one half of the so-called
``effective volume,'' 
\begin{equation}
w(k,\mu)\equiv \frac12\left[\frac{n_gP_g(k,\mu)}{1+n_gP_g(k,\mu)}\right]
V_{survey}
\equiv \frac12V_{eff}(k,\mu).
\label{eq:weight_fm}
\end{equation}
The effective volume is equal to the actual survey volume, $V_{survey}$,
in the sampling variance dominated regime, $P_g(k,\mu)\gg 1/n_g$, whereas it
is small in the shot-noise dominated regime, $P_g(k,\mu)\ll 1/n_g$.
In Figure~\ref{fig4} we show $n_gP_g(k,\mu)$ for
$N_g=0.755\times10^6$ and $b_1=2.5$ as a function of $z$.
The factor of $1/2$ accounts for symmetry in ${\mathbf k}\rightarrow
-{\mathbf k}$.
The derivatives of $\ln P_g(k,\mu)$ with respect to $\theta_i$ are
calculated and given in the Appendix~\ref{sec:derivatives}.

\begin{deluxetable*}{cccccccc}
\tabletypesize{\small}
\tablecaption{Fisher matrix forecast for the full power spectrum
 analysis with various choices of marginalization}
\tablenum{1}
\tablehead{ & \colhead{none} & \colhead{$\ln A$} & \colhead{$\beta$} & \colhead{$\tilde{\sigma}^2_v$} & \colhead{$n_s$} & \colhead{$\alpha_s$} & \colhead{$\ln A$, $\beta$}}
\startdata
$\Delta\ln D_A$ (\%) & 0.279 & 0.877 & 0.317 & 0.282 & 0.479 & 0.416 & 1.100 \\
$\Delta\ln H$ (\%) & 0.437 & 0.786 & 1.124 & 0.801 & 0.509 & 0.539 & 1.134 \\
$r_{D_A,H}$ & 0.382 & $-0.720$ & $-0.309$ & 0.082 & $-0.227$ & $-0.226$
 & 0.038 \\
$\Delta\ln R$ (\%) & 0.187 & 0.762 & 0.317 & 0.259 & 0.386 & 0.363 & 0.775 \\
\hline
 & $\beta$, $\tilde{\sigma}^2_v$ & $\ln A$, $\tilde{\sigma}^2_v$ &  $\ln A$, $\beta$, $\tilde{\sigma}^2_v$ & $\ln A$, $n_s$, $\alpha_s$ & $\ln A$, $\beta$, $n_s$ & $\ln A$, $\beta$, $\tilde{\sigma}^2_v$, \\
& & & & & $\alpha_s$ & $n_s$, $\alpha_s$\\
\hline
$\Delta\ln D_A$ (\%) & 0.327 & 0.891 & 1.101 & 1.089 & 1.233 & 1.250 \\
$\Delta\ln H$ (\%) & 1.457 & 1.101 & 1.468 & 0.984 & 1.362 & 1.530 \\
$r_{D_A,H}$ & $-0.383$ & $-0.632$ & 0.005 & $-0.820$ & $-0.199$ &
 $-0.098$ \\
$\Delta\ln R$ (\%) & 0.322 & 0.869 & 0.879 & 0.974 & 1.000 & 1.014
\enddata
\tablecomments{The fractional errors in $D_A$ and $H$, and their
 cross-correlation coefficients, $r_{D_A,H}$, and the fractional errors
 in the combined 1-d distance scale, $R$ (Eq.~(\ref{eq:sigmaR2})), marginalized over several combinations of parameters:
$\ln A$, $\beta$, $\tilde{\sigma}^2_v$, $\alpha_s$ and $n_s$. The cosmological
parameters are taken from Table 1 of the \citet{komatsu/etal:prep}
 (``WMAP+BAO+SN ML''). 
The survey parameters approximate those of HETDEX:
the survey area and target redshift are $420~{\rm deg^2}$ and
 $1.9<z<3.5$, respectively, 
the number of galaxies is $N_g=0.755\times 10^6$, and the bias
is assumed to be linear with $b_1=2.5$.}
\label{tb:fisher}
\end{deluxetable*}

Unlike for BAOs, which are insensitive to the parameters that affect the
overall shape, for the full modeling we need to make sure that we take
into account potential degeneracy between $D_A$ and $H$ and any other
parameters that affect the overall shape. In this paper we include
$\ln A$, $\beta$, $\tilde{\sigma_v}^2$, $n_s$, and $\alpha_s$.
(We shall comment on the effects of non-linear bias in
\S~\ref{sec:caveat}).

We study the effects of marginalization over various parameter
combinations by taking the submatrix, $\bar{F}_{ij}$,
of the full $7\times 7$  matrix with the index, $i$, of $\theta_i$ running from
1 to 7, such that the submatrix includes the matrix components of desired
parameters to be marginalized. In other words, the parameters that are
not included in the submatrix are fixed and not marginalized over.

Then, we compute the marginalized errors in $\ln D_A$ and $\ln H$ as
\begin{eqnarray}
 \sigma_{\ln D_A} &=& \sqrt{(\bar{F}^{-1})_{11}},\\
 \sigma_{\ln H} &=& \sqrt{(\bar{F}^{-1})_{22}}.
\end{eqnarray}

To simplify the analysis, 
we fix all the other cosmological parameters, such as
$f(z)$, $\Omega_bh^2$, etc. These cosmological parameters will be
determined by the future CMB mission, Planck, accurately, and therefore
it is a good approximation to simply fix them, and vary only $\ln D_A$,
$\ln H$, $A$, $\beta$, $\tilde{\sigma}^2_v$, $n_s$ and $\alpha_s$. 
The fiducial value for the bias is set to $b_1=2.5$ and $f=d\ln D/d\ln a$ is
computed from the fiducial cosmological model.\footnote{One might also
wish to marginalize over $f$ for the following reason: while $f$ can be
calculated from the cosmological parameters assuming the validity of
General Relativity, one might choose to let $f$ free and use it 
for testing  the validity of General Relativity. In this paper we chose
to assume the validity of General Relativity, but one can extend our
analysis to let $f$ free in a straightforward manner.}
We expect that
the analysis of the bispectrum (Fourier transform of three-point
function) will give a precise determination of $b_1$ (as well as
non-linear bias parameters such as $b_2$)
\citep{sefusatti/komatsu:2007}, and therefore it is also a good
approximation to simply fix it. However, 
we also explore a more conservative case where we do not know what $b_1$
is, i.e., we marginalize over the overall amplitude as well as $\beta$. 
In the future work we also plan to
investigate the effect of marginalization over $b_2$, using a joint
analysis of the power spectrum and bispectrum. 
Therefore, our calculation
presented here will provide the lower limit to the errors in 
$\ln D_A$ and $\ln H$ expected from the full modeling of the power spectrum
measured in a survey like HETDEX.
We use the same survey parameters that we have used in
\S~\ref{sec:err_fitex2d}, and we integrate Eq.~(\ref{eq:fisher_fm}) up to
$k_{max}=0.40~h~{\rm Mpc}^{-1}$.

In Figure~\ref{fig3} we show the resulting
error ellipses from the full modeling, in the smaller, solid contours,
with four choices of marginalization. 
(We present the results from more choices of marginalization in
Table~\ref{tb:fisher}.) 
First, for all choices of marginalization we find that the sizes of
the errors in both $D_A$ and $H$ are substantially smaller than those from
the BAO-only analysis with the FITEX-2d.
For example, determinations of both $D_A$
and $H$ are  improved by more than a factor of two in the case of the
amplitude marginalization. 
This is expected, as we are
able to use more information encoded in the power spectrum; namely, the
Hubble horizon at the matter-radiation equality epoch and the Silk
damping scale.  Second, $D_A$ and $H$ are {\it
anti-correlated} for the amplitude marginalization,
with the cross-correlation coefficient of $r=-0.72$ (see top-left panel
of Fig.~\ref{fig3}),
as opposed to a positive correlation seen in the
BAO-only analysis. This is due to the marginalization over the overall
amplitude: if we fixed the overall normalization, then we would still
find a positive correlation between $D_A$ and $H$ with $r=0.38$.

The origin of the negative correlation is the so-called 
Alcock-Paczynski (AP) test \citep{alcock/paczynski:1979}:
when the redshift space distortion is known perfectly well, 
the departure of the power spectrum in redshift space from isotropy,
i.e., dependence of $P(k,\mu)$ on $\mu^2$, can be used to determine
$D_AH$, resulting in $r=-1$ for a power-law power spectrum. 
The contributions from departures of $P(k)$ from a pure power-law,
i.e., the existence of ``standard rulers,'' such as BAOs, the Hubble
horizon at the matter-radiation equality and the Silk damping scale,
make $r$ bigger than $-1$. (See Appendix~\ref{sec:cc} for more details.)
When $\ln A$ and $\beta$ are marginalized over simultaneously, the
correlation between $D_A$ and $H$ 
nearly disappears: the AP test no longer works when we
marginalize over the linear redshift space distortion.
We find $r=0.038$ (see top-right panel of
Fig.~\ref{fig3}) for this case.

When $\ln A$ is marginalized over while the other parameters 
($\beta$, $\tilde{\sigma}_v^2$, $n_s$, and $\alpha_s$) are held fixed, 
we find 0.88\% and 0.79\% errors on
$D_A$ and $H$, respectively, with $r=-0.72$.
The more parameters we marginalize over, the greater 
the cross-correlation coefficient between $D_A$ and $H$ 
as well as the errors on $D_A$ and $H$ become. 
Note that the increase in the errors does not necessarily imply the
decrease in the statistical power in constraining dark energy
properties: since the cross-correlation coefficient is also reduced, the
error in the combined 1-d distance scale, $R$, is much less affected by
the marginalization (see Table~\ref{tb:fisher}). The error in $\ln R$
has been computed as \citep{seo/eisenstein:2007}:
\begin{equation}
 \sigma_{\ln R}^2
= \frac{\sigma_{\ln D_A}^2\left(1-r^2\right)}{1+2r\sigma_{\ln
D_A}/\sigma_{\ln H}+\sigma_{\ln D_A}^2/\sigma_{\ln H}^2}.
\label{eq:sigmaR2}
\end{equation}

Finally, the errors in $D_A$, $H$, and $R$ for various choices of 
marginalization are:
($\sigma_{\ln D_A},\sigma_{\ln H},\sigma_{\ln R}$) = (0.88\%, 0.79\%, 0.76\%),
(1.10\%, 1.13\%, 0.78\%),
(1.10\%, 1.47\%, 0.88\%), and (1.25\%, 1.53\%, 1.01\%) for the marginalization
over $\ln A$, 
$\ln A$ and $\beta$, $\ln A$, $\beta$ and $\tilde{\sigma}^2_v$, and
$\ln A$, $\beta$, $\tilde{\sigma}^2_v$, $n_s$ and $\alpha_s$,
respectively
(see Table~\ref{tb:fisher} for more comprehensive list).
This result should be compared with that from the BAO-only analysis:
($\sigma_{\ln D_A},\sigma_{\ln H},\sigma_{\ln R}$)=(1.76\%, 2.47\%, 1.08\%).
It is clear that the full analysis, even with a generous set of
marginalization choices, beats the BAO-only analysis with a significant
gain in the distance determination accuracies.

\subsection{Caveat for the full modeling}
\label{sec:caveat}
Our analysis presented in \S~\ref{sec:err_full} is too simplistic and 
optimistic, as it ignores any
systematic errors due to our lack of understanding of the effects of
various non-linearities in the power spectrum.

Among the three major
non-linearities, non-linear matter clustering is under control, at least
for high redshifts, i.e., $z\gtrsim 2$, as one can model non-linear
evolution of matter fluctuations almost exactly by the 3rd-order
perturbation theory \citep{jeong/komatsu:2006}. 
While the nominal 3rd-order perturbation theory breaks down at lower
redshifts, $z\sim 1$, there have been a number of studies aiming at
improving upon our ability to compute $P(k)$ at $z\sim 1$ or even lower
redshifts 
\citep{crocce/scoccimarro:2008,matarrese/pietroni:2007,taruya/hiramatsu:2008,valageas:2007,matsubara:2008,mcdonald:2007}. Therefore,
it is quite possible that the non-linear matter clustering will be fully
under control in the near future, at least for the scales that are
relevant to the BAO scales, i.e., $k\lesssim 0.40~h~{\rm Mpc}^{-1}$.

In a separate paper \citep{jeong/komatsu:2008}, we show that
non-linear galaxy biasing is also under control in the weakly non-linear
regime. One can use the perturbation theory approach combined with the
local bias assumption \citep{fry/gaztanaga:1993,mcdonald:2006} to model
the galaxy power spectrum with non-linear bias.

The most problematic one is the non-linear redshift space distortion.
Our understanding of non-linear redshift space distortion, especially
the Finger-of-God (FoG) effect, is limited \citep{scoccimarro:2004}.
Therefore, whether one can achieve the accuracy of $D_A$ and $H$ ($H$ in
particular) reported in Fig.~\ref{fig3} depends crucially on our ability
to correct for the FoG effect. This is work in progress.
Note that the marginalization over $\tilde{\sigma}_v^2$ should capture
some of the increase in the errors in distance scales due to our
ignorance of FoG.

\section{Extraction of $D_A$ and $H$ from the Millennium Simulation}
\label{sec:millennium}
How realistic is our result for the determinations of $D_A$ and $H$ from
the BAO phases using the FITEX-2d method? 
Since our Monte Carlo simulations used in \S~\ref{sec:err_fitex2d} are
too simple, in this section we test the FITEX-2d method further by using
the Millennium Simulation \citep{springel/etal:2005}. 

We use the Millennium Galaxy catalogue, generated by the semi-analytical
galaxy formation code \citep{bower/etal:2006,benson/etal:2003,cole/etal:2000}. 
We have measured the two-dimensional power spectrum of galaxies in
redshift space from the
Millennium Simulation, and applied the FITEX-2d method to remove the
smooth component. We then find the best-fitting $D_A$ and $H$ from the
BAO phases extracted from the FITEX-2d. Again, we use the data up to
$k_{max}=0.40~h~{\rm Mpc}^{-1}$.

\begin{figure}[t]
\centering
\rotatebox{0}{%
  \includegraphics[width=7.5cm]{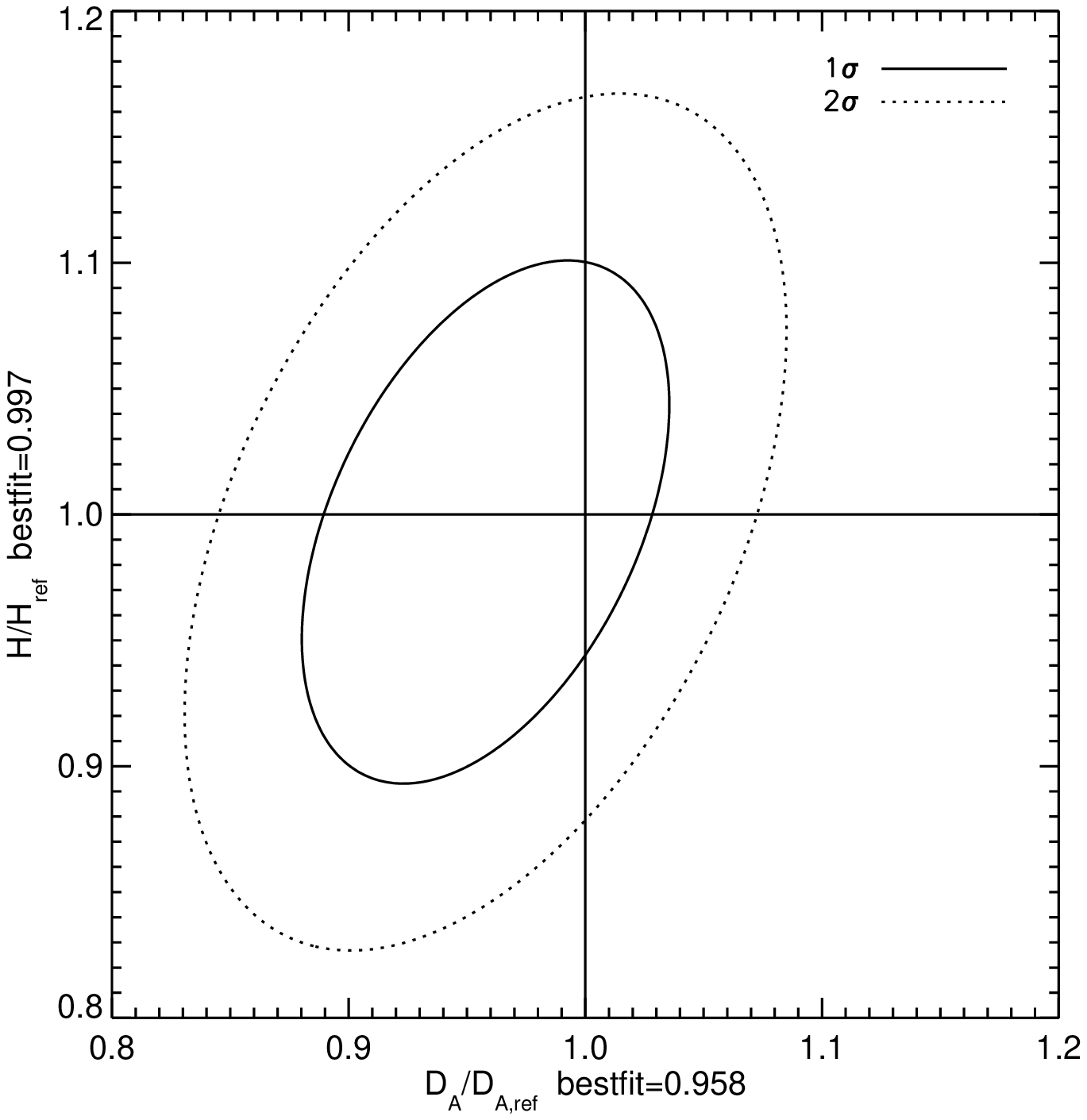}
}%
\caption{%
Accuracy of $D_A$ and $H$ extracted from BAOs with the FITEX-2d method
 applied to the Millennium Galaxy Simulation in redshift space at $z=3$
 \citep{springel/etal:2005,bower/etal:2006,benson/etal:2003,cole/etal:2000}. 
The best-fitting values of $D_A$ and $H$ agree with the true values to
 within statistical errors of the Millennium Simulation;  thus, the
 FITEX-2d method also yields unbiased estimates of $D_A$ and $H$ for the
 Millennium Simulation. The solid and dotted lines show $68\%$ and
 $95\%$ C.L., respectively. 
}%
\label{fig5}
\end{figure}

In Figure~\ref{fig5} we show the result. The best-fitting value that we
find from the Millennium Simulation corresponds to one point at the
center of the contours. We find the errors from the Monte Carlo
simulations that we described in \S~\ref{sec:err_fitex2d} with the survey
parameters replaced by those of the Millennium Simulation:
$V_{survey}= (0.5~h^{-1}~{\rm Gpc})^3$, $n_g=0.138~h^3~{\rm Mpc}^{-3}$,
and $z=3.06$. (There are 17,238,935 galaxies in the Millennium
Simulation at $z=3.06$.) For the theoretical power spectrum that we use
for generating Monte Carlo simulations, we use the best-fitting power
spectrum for the galaxy catalogue of the Millennium Simulation
found in \citet{jeong/komatsu:2008}.

Since the volume of the Millennium Simulation is $\sim\!24$ times as small as
that would be surveyed by HETDEX, the uncertainties in $D_A$ and $H$ are
larger for the Millennium Simulation. (Compare Fig.~\ref{fig5} with
the larger contours of Fig.~\ref{fig3}.)
We find 5.1\% and 6.8\% errors on
$D_A$ and $H$, respectively, with the cross-correlation coefficient of
$r=0.43$, from the Monte Carlo simulations. These errors are larger than 
those from HETDEX Monte-Carlo simulation
by a factor of two (rather than $\sqrt{24}\sim 5$) as the shot noise on the 
power spectrum of the Millennium Simulation is much smaller than that of 
HETDEX simulation.

The best-fitting values of $D_A$ and $H$ are well within
68\% C.L. region, which indicates that the FITEX-2d is able to yield
unbiased estimates of the BAO phases from the Millennium Simulation.

These results indicate that the FITEX-2d method that we have developed
in this paper can be used for extracting the BAOs and measuring $D_A$
and $H$ safely from the real data. It would be interesting to apply the
FITEX-2d method to the two-dimensional power spectrum measured from the
SDSS LRG samples \citep{okumura/etal:prep}, and extract $D_A$ and $H$
from them.

\section{Error Propagation to The Dark Energy Equation of State}
\label{sec:darkenergy}

In \S~\ref{sec:montecarlo} and \S~\ref{sec:millennium}, we have estimated 
errors in $D_A$ and $H$ from two different approaches, i.e., the BAO
fitting using the FITEX-2d method and the full modeling. In this section, we 
propagate errors in $D_A$ and $H$ to those in the dark energy equation
of state parameters.
We parametrize $w(z)$ using the linear model, $w(z)=w_0+w_az/(1+z)$
\citep{linder:2003,chevallier/polarski:2001}.

We propagate the errors in   $D_A$ and $H$ to those in $w_0$ and $w_a$ by
\begin{equation}
\tilde{F}_{\alpha\beta}=\sum_{ij}{\partial p_i\over \partial q_\alpha}{\partial p_j\over \partial q_\beta}F_{ij},
\end{equation}
where $\tilde{F}_{\alpha\beta}$ is the Fisher matrix for the dark energy
parameters, $F_{ij}$ is the Fisher matrix for $D_A$ and $H$, 
$p_i=(\ln D_A, \ln H)$ for $i=1$ and 2, and $q_\alpha=(w_0,w_a)$ for
$\alpha=1$ and 2. 

Partial derivatives of $D_A$ and $H$ with respect to $w_0$ and 
$w_a$ are given by
\begin{eqnarray}
{\partial \ln{D_A} \over \partial w_0}
&=&-{3\over2}\Omega_\Lambda{\int_0^z \ln(1+z^{\prime})f(z^{\prime})g(z^{\prime})^{-3/2}dz^{\prime}
\over\int_0^z g(z^{\prime})^{-1/2}dz^{\prime}},\\
{\partial \ln{D_A} \over \partial w_a}&=&-{3\over2}\Omega_\Lambda\nonumber\\
&\times&{\int_0^z [\ln(1+z^{\prime})-\frac{z^{\prime}}{1+z^{\prime}}]f(z^{\prime})g(z^{\prime})^{-3/2}dz^{\prime}
\over\int_0^z g(z^{\prime})^{-1/2}dz^{\prime}},\\
{\partial \ln{H} \over \partial w_0}&=&{3\over2}\Omega_\Lambda \ln(1+z){f(z)\over g(z)},\\
{\partial \ln{H} \over \partial w_a}&=&{3\over2}\Omega_\Lambda
\left[\ln(1+z)-\frac{z}{1+z}\right]{f(z)\over g(z)},
\end{eqnarray}
where $f(z)$ and $g(z)$ are given by
\begin{eqnarray}
f(z)&=&\exp\left(3\int_0^{z}{1+w_0+w_a \frac{z^{\prime}}{1+z'}
\over 1+z^{\prime}}dz^{\prime}\right),\\
g(z)&=&\Omega_m(1+z)^3+\Omega_\Lambda f(z).
\end{eqnarray}

\begin{figure}[t]
\centering
\rotatebox{0}{%
  \includegraphics[width=9cm]{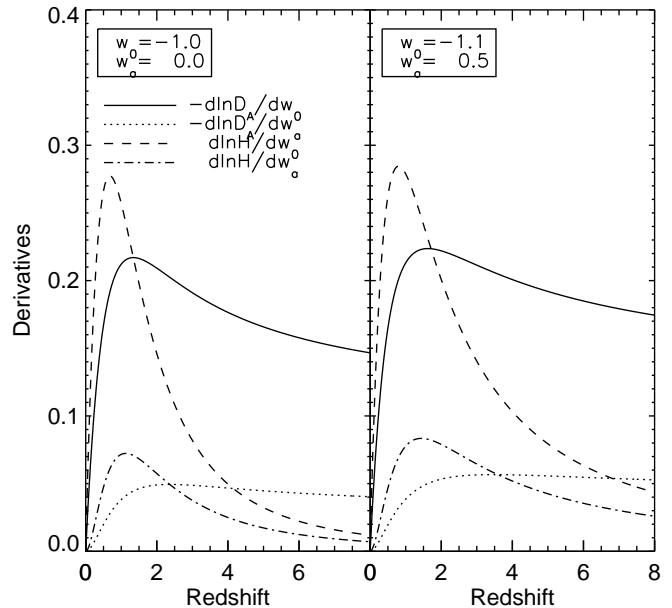}
}%
\caption{%
Partial derivatives of $\ln D_A$ and $\ln H$ with respect to the 
dark energy equation of state parameters, $w_0$ and $w_a$, as a function
 of $z$ for two different cosmological models.
({\it Left}) 
$(w_0, w_a)=(-1.0, 0.0)$.
({\it Right}) 
$(w_0, w_a)=(-1.1, 0.5)$.
}%
\label{fig6}
\end{figure}

Figure \ref{fig6} shows the derivatives as a function of $z$ between
$0.5\le z\le 6.5$ in 
two different cosmological models, ($w_0$, $w_a$)=($-1.0$, $0.0$) and 
($-1.1$, $0.5$). The former is the $\Lambda$CDM model, while the latter
resembles the maximum likelihood values of $w_0$ and $w_a$ from the 
WMAP+BAO+SN+BBN \citep{komatsu/etal:prep}.
The derivatives are similar for these cosmological models, and therefore
we use the $\Lambda$CDM model as the fiducial model for computing the
derivatives. 

We add the distance information from CMB as
\begin{equation}
\tilde{F}^{total}_{\alpha\beta}(z)=\tilde{F}^{CMB}_{\alpha\beta}
+\tilde{F}^{gal}_{\alpha\beta}(z),
\end{equation}
where we assume that the CMB experiment yields 1\% determination of the
angular diameter distance out to $z=1090$, i.e., we use
\begin{equation}
 \tilde{F}^{CMB}_{\alpha\beta} = 10^4 \frac{\partial \ln
  D_A(z=1090)}{\partial q_\alpha}\frac{\partial \ln D_A(z=1090)}{\partial q_\beta}.
\end{equation}

\begin{figure}[t]
\centering
\rotatebox{0}{%
  \includegraphics[width=9cm]{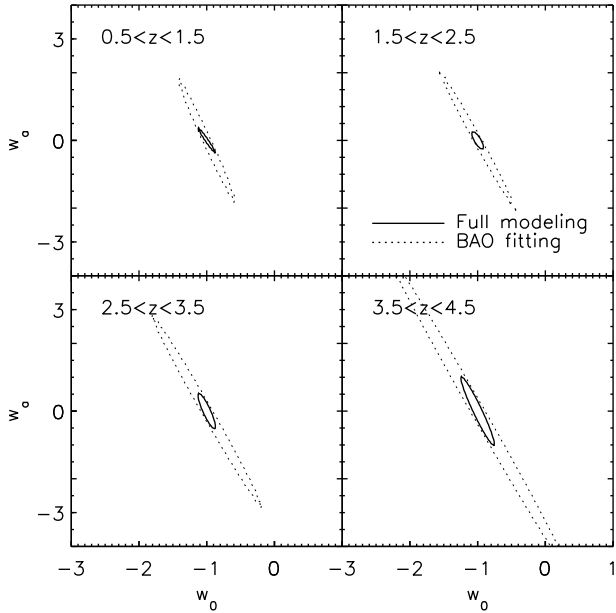}
}%
\caption{%
Projected 68\% constraints on the dark energy parameters, $w_0$ and
 $w_a$: the BAO fitting with the FITEX-2d method (dotted) versus the
 full modeling (solid). For both cases, we use the power spectrum 
up to $k_{max}=0.40~h~{\rm Mpc}^{-1}$, and we assume that the CMB
 experiment measures the angular diameter distance out to $z=1090$ with
 1\% accuracy. The survey area and the number of galaxies are
 420~${\rm deg^2}$ and $N_g=0.755\times10^6$ for all cases.
({\it Top Left}) $0.5\le z\le 1.5$,
({\it Top Right}) $1.5\le z\le 2.5$,
({\it Bottom Left}) $2.5\le z\le 3.5$,
({\it Bottom Right}) $3.5\le z\le 4.5$.
}%
\label{fig7}
\end{figure}

We are interested in how the BAO-only analysis compares with the full
modeling. In Fig.~\ref{fig7} we show the projected error contours on
$w_0$ and $w_a$ calculated from the BAO-only analysis with the FITEX-2d
and those from the full analysis at four redshift bins:
$0.5\le z\le 1.5$, $1.5\le z\le 2.5$, $2.5\le z\le 3.5$, and $3.5\le
z\le 4.5$. The survey area and the number of galaxies are
 420~${\rm deg^2}$ and $N_g=2.9\times10^6$ for all redshift bins.
From the BAO-only analysis we find $(\Delta w_0,\Delta w_a)=(0.29,1.26)$, 
$(0.38,1.39)$, $(0.55,1.92)$, and $(0.91,3.18)$, whereas
from the full modeling we find $(\Delta w_0,\Delta w_a)=(0.09,0.27)$, 
$(0.06,0.17)$, $(0.09,0.35)$, and $(0.17,0.68)$, for
$0.5\le z\le 1.5$, $1.5\le z\le 2.5$, $2.5\le z\le 3.5$, and $3.5\le
z\le 4.5$, respectively.

We therefore conclude that the full analysis yields much better
constraints on $w_0$ and $w_a$ than the BAO-only analysis.

\section{Conclusion}
\label{sec:conclusion}
In this paper we have developed a method, called the FITEX-2d method, to
extract the two-dimensional phases of BAOs from galaxy power spectra in
redshift space. Our model builds on and extends the existing
one-dimensional algorithm, called FITEX, developed by
\citet{koehler/schuecker/gebhardt:2007}. 

Our method removes the smooth, non-oscillating component from the
observed galaxy power spectrum in redshift space. The fitting function
consists of the smooth one-dimensional spectrum that depends only on
$k$, $P_{smooth}^{1d}(k)$ given by Eq.~(\ref{eq:Psmooth1d}), multiplied by 
the angle-dependent function expanded in the Legendre polynomials with
even multipoles. The coefficients of the Legendre polynomials contain
even powers of $k$. 
The resulting function, given by Eq.~(\ref{eq:Psmooth2d}), is able to capture
the non-oscillating part of the galaxy power spectrum well.

We have tested the FITEX-2d method using the analytical model without
any noise, the Monte Carlo realizations with noise expected from the
HETDEX experiment \citep{hill/etal:2004}, and the galaxy catalogue
created from the Millennium Simulation \citep{springel/etal:2005}.
In all cases the FITEX-2d method yields unbiased estimates of the
angular diameter distance, $D_A$, and the expansion rate, $H$. 

However, the BAOs capture only a part of distance information encoded in the
galaxy power spectrum. To exploit the distance information, especially
the equality scale, $r_H(z_{eq})$, we have explored the constraints on
$D_A$ and $H$ from the full modeling of the galaxy power spectrum in
redshift space. Provided that three key non-linearities (non-linear
matter clustering, non-linear galaxy bias, and non-linear redshift space
distortion) are under control, we find that the full modeling yields the
constraints that are better than the BAO-only analysis by more than a
factor of two both in $D_A$ and $H$, and the dark energy parameters such
as $w_0$  and $w_a$. 

While the effects of non-linear matter clustering
\citep{jeong/komatsu:2006,crocce/scoccimarro:2008,matarrese/pietroni:2007,taruya/hiramatsu:2008,valageas:2007,matsubara:2008,mcdonald:2007}
and non-linear galaxy bias \citep{jeong/komatsu:2008} are being
understood in the weakly non-linear regime that is relevant to the
future galaxy surveys, the effects of non-linear redshift space
distortion are poorly understood. While the FITEX-2d method that we have
developed in this paper are useful for obtaining {\it robust} constraints
on $D_A$ and $H$, hence the dark energy properties, one must understand
non-linear redshift space distortion to fully exploit the full
information content of the galaxy power spectrum in redshift space.
We would then be able to reduce the errors in $D_A$ and $H$ by more than
a factor of two.

\acknowledgements
We thank the anonymous referee for illuminating comments on the
correlation coefficients, which motivated our doing more thorough analysis.
This material is based in part upon work supported by the Texas Advanced
Research Program under Grant No. 003658-0005-2006.
EK acknowledges support from an Alfred P. Sloan Research Fellowship.
The Millennium Simulation databases used in this paper and the 
web application providing online access to them were constructed as 
part of the activities of the German Astrophysical Virtual Observatory.
\appendix
\section{Fisher Matrix Code}
\label{sec:cc_fisher}
In this Appendix we describe what we have implemented in our Fisher
matrix code, which is publicly available as a part of ``Cosmology
Routine Library (CRL),'' developed by one of the authors (EK).
This code includes the non-linear matter power spectrum in both real and
redshift space, as well as marginalization over the amplitude, the
linear redshift space distortion, the velocity dispersion of
Fingers-of-God, the primordial tilt and running index. In the future
release we plan to include non-linear galaxy bias and primordial
non-Gaussianity. 

\subsection{Basics}
A simple, approximate formula of the Fisher matrix for galaxy survey
is given by \citep[e.g.,][]{seo/eisenstein:2003}
\begin{equation}
F_{ij}=\int_0^{k_{max}}\frac{4\pi k^2dk}{(2\pi)^3}
\int_0^1d\mu\frac{\partial\ln P_g(k,\mu)}{\partial\theta_i}
\frac{\partial\ln P_g(k,\mu)}{\partial\theta_j}w(k,\mu),
\label{eq:fisher}
\end{equation}
where $P_g(k,\mu)$ is the galaxy survey power spectrum calculated
theoretically as a function of parameters, $\theta_i$ are the
parameters to be extracted from the data, and $w(k,\mu)$
is a function given by
\begin{equation}
w(k,\mu)\equiv \frac12\left[\frac{n_gP_g(k,\mu)}{1+n_gP_g(k,\mu)}\right]
V_{survey}
\label{eq:weight}
\end{equation}
Here, $n_g$ and $V_{survey}$ are the number density of galaxies and
the volume of survey, respectively. 

In linear theory, $P_g(k,\mu)$ is given by
\begin{equation}
P_g(k,\mu)=b_1^2R(\mu^2)P^{linear}(k),
\end{equation}
where $b_1$ is the scale independent linear bias factor, $P^{linear}(k)$
is the linear matter power spectrum, and $R(\mu^2)$ describes the
linear redshift space distortion effect (Kaiser effect):
\begin{eqnarray}
R(\mu^2)&\equiv& (1+\beta\mu^2)^2\\
\beta&=&(d\ln D/d\ln a)/b_1,
\end{eqnarray}
where $D$ is the growth factor of the linear density fluctuations,
and $a$ is the scale factor.
\subsection{Derivatives}
\label{sec:derivatives}
To calculate the logarithmic derivatives of $P(k)$ in Eq.~(\ref{eq:fisher}), 
let us write down the non-linear galaxy power spectrum (with linear bias)
as (Eq.~(\ref{eq:pkred})):
\begin{eqnarray}
\nonumber
 P_g(k,\mu) = b_1^2\left[P_{\delta\delta}(k) + 2\beta\mu^2P_{\delta\theta}(k)
+\beta^2\mu^4P_{\theta\theta}(k)\right]
\times\frac1{1+f^2k^2\mu^2\tilde{\sigma}_v^2}.
\end{eqnarray}
We compute the derivatives with respect to the following seven
parameters: 
the angular diameter distance, $D_A$, the Hubble expansion rate, $H$,
the overall amplitude of the galaxy power spectrum, $A$,  the linear
redshift space distortion factor, $\beta\equiv f/b_1$, the velocity
dispersion with an empirically calibrated fudge factor, $\tilde{\sigma}^2_v$,
the tilt of the primordial power spectrum, $n_s$, and the running index,
$\alpha_s$ ($P_{ini}\propto k^{n_s+1/2\alpha_s\ln[k/k_{pivot}]}$).
We choose the convention such that
\begin{equation}
(\theta_1,\theta_2,\theta_3,\theta_4,\theta_5,\theta_6,\theta_7)
=(\ln D_A,\ln H,\ln A,\beta,\tilde{\sigma}^2_v,n_s,\alpha_s).
\end{equation}

The derivatives with respect to $\ln A$, $\beta$, $\tilde{\sigma}_v^2$,
$n_s$, and $\alpha_s$ are easy to evaluate. They are given by

\begin{eqnarray}
\frac{\partial\ln P_g(k,\mu)}{\partial\ln A}&=&1,\\
\frac{\partial\ln P_g(k,\mu)}{\partial\beta}&=&
\frac{2\mu^2P_{\delta\theta}(k)+2\beta\mu^4P_{\theta\theta}(k)}
{P_{\delta\delta}(k)+2\beta\mu^2P_{\delta\theta}(k)+\beta^2\mu^4P_{\theta\theta}(k)}
\label{eq:dlnpdbeta}
\\,
\frac{\partial\ln P_g(k,\mu)}{\partial\tilde{\sigma}^2_v}&=&
-\frac{f^2k^2\mu^2}{1+f^2k^2\mu^2\tilde{\sigma}^2_v},\\
\frac{\partial\ln P_g(k,\mu)}{\partial n_s}&=&
\frac{\partial\ln P_{ini}(k)}{\partial n_s}=\ln k,\\
\frac{\partial\ln P_g(k,\mu)}{\partial\alpha_s}&=&
\frac{\partial\ln P_{ini}(k)}{\partial \alpha_s}=
\frac12\left[\ln\left(\frac{k}{k_{pivot}}\right)\right]^2.
\end{eqnarray}
We compute the derivatives with respect to $\ln D_A$ and $\ln H$
in a two step process. First, we write
\begin{eqnarray}
\frac{\partial\ln P_g(k,\mu)}{\partial\ln D_A}&=&
\frac{\partial\ln P_g(k,\mu)}{\partial\ln k}
\frac{\partial\ln k}{\partial\ln D_A}
+\frac{\partial\ln P_g(k,\mu)}{\partial\mu^2}
\frac{\partial\mu^2}{\partial\ln D_A},\\
\frac{\partial\ln P_g(k,\mu)}{\partial\ln H}&=&
\frac{\partial\ln P_g(k,\mu)}{\partial\ln k}
\frac{\partial\ln k}{\partial\ln H}
+\frac{\partial\ln P_g(k,\mu)}{\partial\mu^2}
\frac{\partial\mu^2}{\partial\ln H},
\end{eqnarray}
where
\begin{eqnarray}
\frac{\partial\ln k}{\partial\ln D_A}&=&1-\mu^2,\\
\frac{\partial\ln k}{\partial\ln H}&=&-\mu^2,\\
\frac{\partial\mu^2}{\partial\ln D_A}&=&-2\mu^2(1-\mu^2),\\
\frac{\partial\mu^2}{\partial\ln H}&=&-2\mu^2(1-\mu^2),\\
\frac{\partial\ln P_g(k,\mu)}{\partial\mu^2}&=&
\frac{2\beta P_{\delta\theta}(k)+2\beta^2\mu^2P_{\theta\theta}(k)}
{P_{\delta\delta}(k)+2\beta\mu^2P_{\delta\theta}(k)+\beta^2\mu^4P_{\theta\theta}(k)}
-\frac{f^2 k^2\tilde{\sigma}_v^2}
{1+f^2k^2\mu^2\tilde{\sigma}_v^2}.
\end{eqnarray}
Finally, we need to know the ``effective spectral index'', $n_{eff}(k,\mu)$,
given by
\begin{equation}
n_{eff}(k,\mu)\equiv\frac{\partial\ln P_g(k,\mu)}{\partial\ln k},
\label{eq:n_eff}
\end{equation}
or explicitly
\begin{eqnarray}
n_{eff}(k,\mu)&=&\frac{P_{\delta\delta}(k)n_{\delta\delta}(k)
+2\beta\mu^2P_{\delta\theta}(k)n_{\delta\theta}(k)
+\beta^2\mu^4P_{\theta\theta}(k)n_{\theta\theta}(k)}
{P_{\delta\delta}(k)+2\beta\mu^2P_{\delta\theta}(k)+\beta^2\mu^4P_{\theta\theta}(k)}\nonumber\\
&&-\frac{2f^2 k^2\mu^2\tilde{\sigma}_v^2}{1+f^2k^2\mu^2\tilde{\sigma}_v^2},
\label{eq:n_eff_exp}
\end{eqnarray}
where
\begin{eqnarray}
n_{\delta\delta}(k)&\equiv&
\frac{\partial\ln P_{\delta\delta}(k)}{\partial\ln k},\\
n_{\delta\theta}(k)&\equiv&
\frac{\partial\ln P_{\delta\theta}(k)}{\partial\ln k},\\
n_{\theta\theta}(k)&\equiv&
\frac{\partial\ln P_{\theta\theta}(k)}{\partial\ln k}.
\end{eqnarray}
\subsection{Correlation Coefficients}
\label{sec:cc}
In this subsection we explore the behaviour of the 
cross-correlation coefficient between $D_A$ and $H$ in various cases. In
particular we focus on the effect of the marginalization over the
overall amplitude with (\S~\ref{sec_cc_beta}) and 
without (\S~\ref{sec:cc_no_beta}) the additional marginalization over
the redshift space distortion.
\subsubsection{No redshift space distortion, $\beta=0$}
\label{sec:cc_no_beta}
Let us evaluate the Fisher matrices in the limit that the redshift
space distortion is absent, i.e., $\beta=0$. In this limit, the
weighting function in Eq.~(\ref{eq:fisher}) and the effective spectral
index in Eq.~(\ref{eq:n_eff}) become independent
of $\mu$, i.e., $w(k,\mu)\to w(k)$ and $n_{eff}(k,\mu)\to n_{eff}(k)$.
We obtain
\begin{eqnarray}
F_{11}&=&\int^{k_{max}}_{k_{min}}\frac{k^2dk}{2\pi^2}[n_{eff}(k)]^2w(k)
\int^1_0d\mu (1-\mu^2)^2\\
F_{12}&=&\int^{k_{max}}_{k_{min}}\frac{k^2dk}{2\pi^2}[n_{eff}(k)]^2w(k)
\int^1_0d\mu (1-\mu^2)(-\mu^2)\\
F_{13}&=&\int^{k_{max}}_{k_{min}}\frac{k^2dk}{2\pi^2}n_{eff}(k)w(k)
\int^1_0d\mu (1-\mu^2)\\
F_{22}&=&\int^{k_{max}}_{k_{min}}\frac{k^2dk}{2\pi^2}[n_{eff}(k)]^2w(k)
\int^1_0d\mu (-\mu^2)^2\\
F_{23}&=&\int^{k_{max}}_{k_{min}}\frac{k^2dk}{2\pi^2}n_{eff}(k)w(k)
\int^1_0d\mu (-\mu^2)\\
F_{33}&=&\int^{k_{max}}_{k_{min}}\frac{k^2dk}{2\pi^2}w(k)
\int^1_0d\mu
\end{eqnarray}
Now, in order to understand the effect of the structure of
$n_{eff}$, let us assume that the galaxy power spectrum is a pure
power-law, i.e., $n_{eff}(k)=n$ and $n$ is the independent of $k$.
In this limit, we obtain

\begin{equation}
F_{ij}=\bar{w}
\left(
\begin{array}{ccc}
\frac{8n^2}{15} & -\frac{2n^2}{15} & \frac{2n}{3}\\
-\frac{2n^2}{15} & \frac{n^2}{5} & -\frac{n}{3}\\
\frac{2n}{3} & -\frac{n}{3} & 1
\end{array}
\right)
\label{eq:3by3fisher}
\end{equation}
where $\bar{w}\equiv\int\frac{k^2dk}{2\pi^2}w(k)$.

The marginalized errors of parameters and the correlation coefficients
are computed from the inverse of the Fisher matrix. However, one can
show that the matrix given in Eq.~(\ref{eq:3by3fisher}) is singular.
In other words, $D_A$ and $H$ are completely degenerate with the amplitude
for a power-law power spectrum. This result shows that only the departure
of the power spectrum from a pure power-law, i.e., the existence
of characteristic scales, can break the degeneracy between $D_A$ and $H$,
and $A$. These scales are often called the ``standard rulers.''

To understand the structure of the Fisher matrix in Eq.~(\ref{eq:3by3fisher})
better, let us add small perturbations, $\epsilon>0$, to the diagonal
elements, and invert the matrix. The result is
\begin{equation}
(F^{-1})_{ij}=\frac1{\bar{w}}
\left(
\begin{array}{ccc}
\frac1{(2+n^2)\epsilon} & -\frac1{(2+n^2)\epsilon} & -\frac{n}{(2+n^2)\epsilon}\\
-\frac1{(2+n^2)\epsilon} & \frac1{(2+n^2)\epsilon} & \frac{n}{(2+n^2)\epsilon}\\
-\frac{n}{(2+n^2)\epsilon} & \frac{n}{(2+n^2)\epsilon} & \frac{n^2}{(2+n^2)\epsilon}
\end{array}
\right)
+\mathcal{O}(\epsilon^0)
\end{equation}
We find that the correlation coefficient between $D_A$ and $H$ is
\begin{equation}
r_{12}\equiv\frac{(F^{-1})_{12}}{\sqrt{(F^{-1})_{11}(F^{-1})_{22}}}
\to -1
\end{equation}
as $\epsilon\to 0$. Therefore, $\ln D_A$ and $\ln H$ are totally
anti-correlated, which implies that, although we cannot determine
$\ln D_A$ and $\ln H$ simultaneously, we can determine
$\ln D_A+\ln H=\ln(D_AH)$, even for a power-law power spectrum.
\footnote{
The other cross-correlation coefficients are $r_{13}\to\mp 1$
and $r_{23}\to\pm 1$ for $n>0$ and $n<0$ respectively.
}
This is known as the Alcock-Paczy\'nski (AP)
test \citep{alcock/paczynski:1979}.

There is a special case in which the covariance between $A$ and $D_A$ or $H$
may be ignored. One may imagine the situation where $n_{eff}(k)$
depends upon $k$ such that $A$ is uncorrelated with $D_A$ or $H$.
For example, if $n_{eff}(k)$ oscillates about zero, then
$\int_{k_{min}}^{k_{max}}\frac{k^2dk}{2\pi^2}n_{eff}(k)w(k)$ would
be small compared with
$\int_{k_{min}}^{k_{max}}\frac{k^2dk}{2\pi^2}[n_{eff}(k)]^2w(k)$
or $\int_{k_{min}}^{k_{max}}\frac{k^2dk}{2\pi^2}w(k)$.
Therefore, $F_{13}$ and $F_{23}$ may be ignored, making $A$ de-correlated
with $D_A$ and $H$. In this case, the Fisher matrix is a 2-by-2 matrix given by
\begin{equation}
F_{ij}=\bar{w}n^2
\left(
\begin{array}{cc}
\frac8{15} & -\frac2{15}\\
-\frac2{15} & \frac15
\end{array}
\right)
\end{equation}
The inverse of this matrix is then
\begin{equation}
(F^{-1})_{ij}=\frac1{\bar{w}n^2}
\left(
\begin{array}{cc}
\frac94 & \frac32\\
\frac32 & 6
\end{array}
\right)
\end{equation}
The correlation coefficient between $D_A$ and $H$ is thus given by
\begin{equation}
r_{12}=\frac{3/2}{\sqrt{9/4\times6}}=\frac1{\sqrt{6}}\simeq0.408.
\end{equation}
This result has been derived by \citet{seo/eisenstein:2007}, and
justifies the use of BAOs as a way to measure $D_A$ and $H$ with a correlation
coefficient of $0.408$.

From these studies we are led to the following conclusion:
\begin{itemize}
\item When the information is dominated by BAOs, the correlation coefficient
between $D_A$ and $H$ is $r_{12}\simeq 0.408$. The amplitude of the
BAOs contributes little to the errors on $D_A$ and $H$, 
as the amplitude information is de-correlated with $D_A$ and $H$.
\item When the information is dominated by the AP test, $r_{12}\simeq -1$.
\item In reality, as we have shown in this paper, BAOs contribute less than the
overall shape of the power spectrum. Also, the shape of the power spectrum
is not exactly a power-law. As a result, the correlation coefficient from the
full analysis is usually negative (or small positive), 
but always greater than $-1$ (see Table~\ref{tb:fisher}).
\end{itemize}
\subsubsection{With redshift space distortion, $\beta>0$}
\label{sec_cc_beta}
Next, let us consider the case where the redshift space distortion
cannot be ignored. In this case, we see from Eq.~(\ref{eq:weight})
and Eq.~(\ref{eq:n_eff_exp}) that
the weighting function, $w(k,\mu)$, and  the effective spectral index,
$n_{eff}(k,\mu)$, are no longer independent of $\mu$.
The analytical treatment is also possible for this case, although the
results are too complicated to be useful. 
We therefore report on the numerical results.

Here, we choose the survey parameters given in \S~\ref{sec:err_full}
with the non-linear power spectrum of Eq.~(\ref{eq:pkred}).
The results from the numerical calculations of the Fisher matrix are 
given in Table \ref{tb:fisher}.
We find that the marginalization over the amplitude information,
and that over the amplitude and the shape of the primordial power
spectrum (i.e., $n_s$ and $\alpha_s$) 
give the cross-correlation close to $-1$; thus, one relies on the
AP test. 
The marginalization over the amplitude and the linear redshift
space distortion (i.e., $\beta$) drive the cross-correlation
towards zero, as the AP test no longer works when the linear redshift
space distortion is marginalized over. However, in both cases the errors
in the combined 1-d distance scale, $R$, are about the same. In other
words, while one changes the orientation of the ellipse, the area is
approximately preserved. 

In summary, when the amplitude information is marginalized over, the information
is mostly coming from the dependence of $P(k,\mu)$ on $\mu^2$, which
yields a constraint on $D_AH$ via the AP test, while when both the amplitude and
the linear redshift space distortion are marginalized, the most information is
coming from the standard rulers, which can constrain $D_A$ and $H$
separately, driving the cross-correlation towards zero.

\begin{figure}[t]
\centering
\rotatebox{0}{%
  \includegraphics[width=9cm]{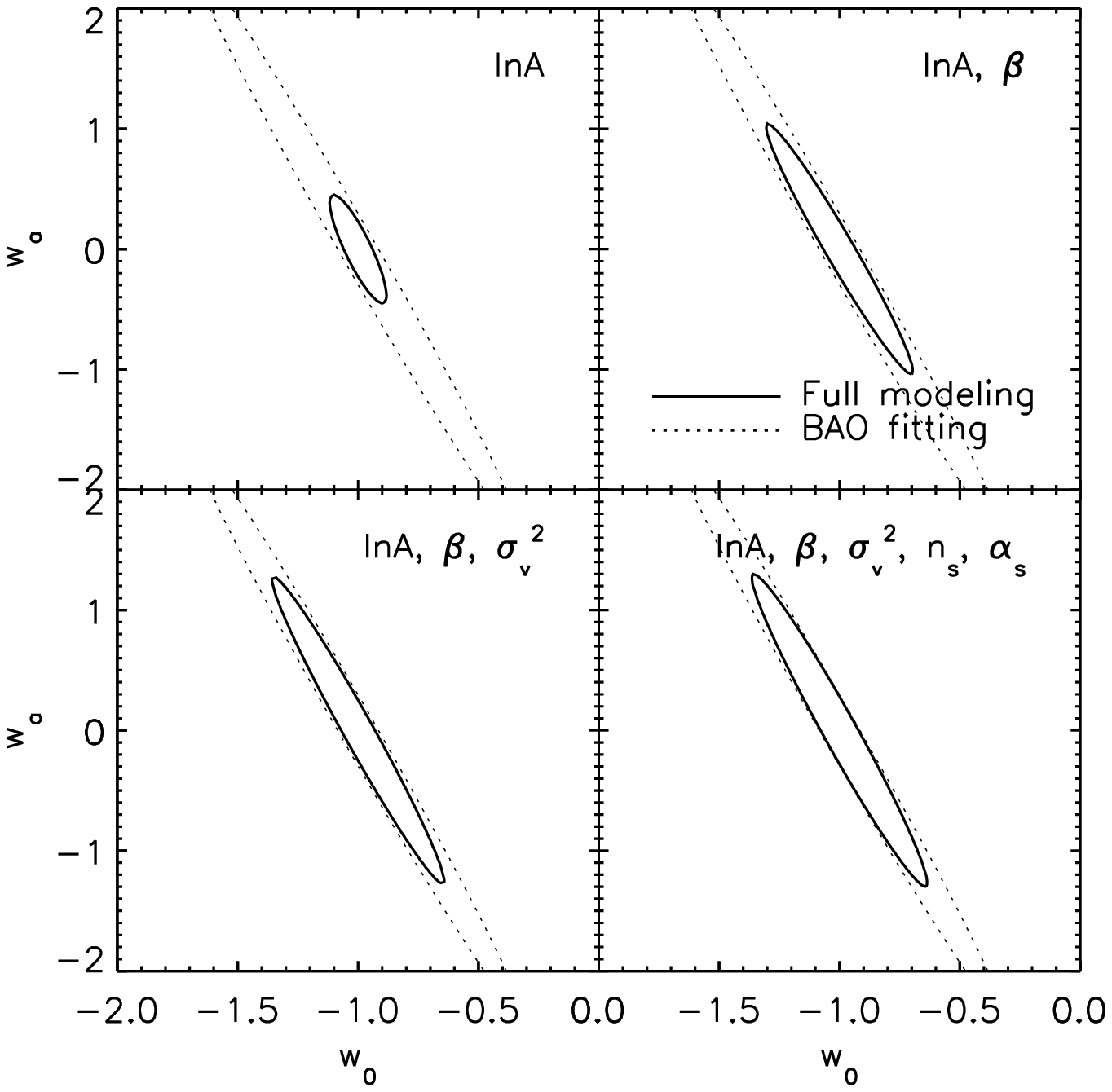}
}%
\caption{%
Projected 68\% constraints on the dark energy parameters, $w_0$ and
$w_a$. The full modeling (solid) marginalized over different combinations of
parameters as well as the BAO-only analysis (dotted) are shown. 
For all cases, we use the power spectrum up to $k_{max}=0.40~h~{\rm Mpc}^{-1}$, and we assume that the CMB
experiment measures the angular diameter distance out to $z=1090$ with
1\% accuracy. The survey area and the number of galaxies are
420~${\rm deg^2}$ and $N_g=0.755\times10^6$, and the redshift range is
$1.9\le z\le 3.5$ for all cases.
({\it Top Left}) marginalized over $\ln A$,
({\it Top Right}) marginalized over $\ln A$, $\beta$,
({\it Bottom Left}) marginalized over $\ln A$, $\beta$ and $\tilde{\sigma}^2_v$,
({\it Bottom Right}) marginalized over $\ln A$, $\beta$, $\tilde{\sigma}^2_v$,
$n_s$ and $\alpha_s$.
}%
\label{fig8}
\end{figure}

Finally, in Fig.~\ref{fig8} we show how different choices of 
marginalization over parameters influence the error contours of $w_0$ and $w_a$:
$(\Delta w_0,\Delta w_a)=(0.08,0.27)$, 
$(0.08,0.30)$, $(0.24,0.85)$, and $(0.24,0.86)$, for the cases of no marginalization,
marginalization over $\ln A$, marginalization over $\ln A$, $\beta$ and $\tilde{\sigma}^2_v$,
and marginalization over $\ln A$, $\beta$, $\tilde{\sigma}^2_v$, $n_s$ and $\alpha_s$
respectively.

\subsection{User's Guide}
When using the Fisher matrix code, one may choose the form of the model
galaxy power spectrum from: 
\begin{itemize}
\item Linear power spectrum with the linear redshift space distortion
      (Kaiser effect), 
\item Non-linear power spectrum from the 3rd-order perturbation theory with
the linear redshift space distortion, 
\item Non-linear power spectrum from the 3rd-order perturbation theory with
the non-linear redshift space distortion given by Eq.~(71) of
      \citet{scoccimarro:2004}, or
\item Non-linear power spectrum from the 3rd-order perturbation theory with
the non-linear redshift space distortion given by Eq.~(\ref{eq:pkred}).
\end{itemize}

Next, specify the number of parameters one wishes to marginalize
over, and then choose the parameters from: $\ln A$, $\beta$,
$\tilde{\sigma}^2_v$, $n_s$, and $\alpha_s$.

A given galaxy survey can be sliced up into multiple redshift bins.
After entering the survey area in units of ${\rm deg}^2$, one is asked
to enter  the following parameters at each redshift bin: 
the redshift range ($z_{min}\!<\!z\!<\!z_{max}$), the number of galaxies
in units of millions in the bin, $b_1$, $k_{max}$ in units of 
$h~{\rm Mpc}^{-1}$, and the redshift error in units of km/s.

The linear power spectrum at $z=30$ has been precomputed using the CAMB code
\citep{lewis/challinor/lasenby:2000} for the
maximum likelihood parameters given in Table~1 of
\citet{komatsu/etal:prep} (''WMAP+BAO+SN''). 
The ingredients of the non-linear power spectra, $P_{\delta\delta}$,
$P_{\delta\theta}$, and $P_{\theta\theta}$, have been precomputed
from the linear spectrum at $z=30$. 
These spectra are then evolved to a specified redshift by the
appropriate growth factor obtained by solving the differential equation
given in Eq.~(76) of \citet{komatsu/etal:prep}.

Finally, the code yields the errors on $\ln D_A$, $\ln H$, $r_{D_A,H}$,
and $\ln R$ (see Eq.~(\ref{eq:sigmaR2}) for the definition of the error
in the combined distance scale, $R$).

\end{document}